\documentclass[reprint,amsmath,amssymb,aps]{revtex4-1}

\usepackage{graphicx} 
\usepackage{braket} 

\begin{document}
	
	\preprint{APS/123-QED} 
	
	\title{Spin-selective AC Stark shifts in a charged quantum dot}
	
	\author{T. A. Wilkinson}
	\affiliation{Department of Physics and Astronomy, West Virginia University, Morgantown, WV 26506, USA}
	
	\author{D. J. Cottrill}
	\affiliation{Department of Physics and Astronomy, West Virginia University, Morgantown, WV 26506, USA}
	
	\author{J. M. Cramlet}
	\affiliation{Department of Physics and Astronomy, West Virginia University, Morgantown, WV 26506, USA}
	
	\author{C. E. Maurer}
	\affiliation{Department of Physics and Astronomy, West Virginia University, Morgantown, WV 26506, USA}
	
	\author{C. J. Flood}
	\affiliation{Department of Physics and Astronomy, West Virginia University, Morgantown, WV 26506, USA}
	
	\author{A. S. Bracker}
	\affiliation{Naval Research Laboratory, Washington, DC 20375}
	
	\author{M. Yakes}
	\affiliation{Naval Research Laboratory, Washington, DC 20375}
	
	\author{D. Gammon}
	\affiliation{Naval Research Laboratory, Washington, DC 20375}
	
	\author{E. B. Flagg}
	\affiliation{Department of Physics and Astronomy, West Virginia University, Morgantown, WV 26506, USA}
	
	\date{\today}

	\begin{abstract}
		\noindent A strong, far-detuned laser can shift the energy levels of an optically active quantum system via the AC Stark effect. We demonstrate that the polarization of the laser results in a spin-selective modification to the energy structure of a charged quantum dot, shifting one spin manifold but not the other. An additional shift occurs due to the Overhauser field of the nuclear spins, which are pumped into a partially polarized state. This mechanism offers a potentially rapid, reversible, and coherent control of the energy structure and polarization selection rules of a charged quantum dot.
	\end{abstract}

	\maketitle 

	%

	Quantum confined spins are regarded as potential candidates for quantum information applications \cite{imamoglu_quantum_1999,liu_quantum_2010,kloeffel_prospects_2013}. The spin eigenstates could act as the two states of a qubit, and they can be manipulated and measured by magnetic and optical fields. In an epitaxially grown quantum dot (QD), the Zeeman splitting due to an external magnetic field is commonly used to control the selection rules of the optical transitions, and lasers can then control the spin in a variety of ways \cite{warburton_single_2013}. Depending on the magnetic field orientation the spin can be initialized \cite{xu_fast_2007,gerardot_optical_2008}, coherently manipulated \cite{press_complete_2008,berezovsky_picosecond_2008,hansom_environment-assisted_2014}, or measured via fluorescence \cite{delteil_observation_2014}. Unfortunately, the magnetic field orientations necessary for manipulation and measurement are orthogonal. Coherent manipulation requires a field transverse to the growth direction (Voigt configuration), while fluorescent measurement requires a field along the growth direction (Faraday configuration). The Voigt configuration lacks cycling transitions that preserve the spin \cite{xu_fast_2007,warburton_single_2013}, which makes single-shot measurement of the spin state nearly impossible. The Faraday configuration lacks allowed optical transitions that link the spin manifolds \cite{warburton_single_2013}, which precludes universal coherent optical manipulation of the spin orientation.
	
	A potential solution for this impasse is to use the AC Stark effect to adjust the energy levels of the QD \cite{muller_resonance_2007}. A strong, circularly polarized, far-detuned laser could apply a spin-selective energy shift. If this shift is significantly larger than the Zeeman splitting in a Voigt configuration, the system will convert to a pseudo-Faraday configuration \cite{flagg_optical_2015}, where the energy structure, eigenstates, and polarization selection rules are similar to those caused by a longitudinal magnetic field. The same reconfiguration could be accomplished in the field due to the local nuclear spins in the QD \cite{hansom_environment-assisted_2014}. The reconfiguration depends on the power of the laser field, so it is reversible and it can be applied or removed rapidly over a few nanoseconds. That would allow switching between manipulation and measurement of an electron spin.
	
	Here we experimentally demonstrate spin-selective AC Stark shifts applied to a charged QD. The transition frequency of one spin manifold shifts by more than 20 GHz, which is much larger than the 1 GHz linewidth, while the other transition is not shifted by the AC Stark effect. The polarization, power, and detuning of the laser causing the AC Stark effect determine the shifts of the transitions. Linear polarization shifts both transitions equally, while circular polarization shifts only one of them. Red-detuning of the laser causes a blue-shift of the transition, while blue-detuning causes a red-shift. In addition to the AC Stark shift, we observe another energy shift caused by dynamic polarization of the nuclear spins of the atoms comprising the QD. The direction of the nuclear polarization is determined by the polarization of the AC Stark laser.
	
	The QDs studied here are self-assembled InGaAs QDs embedded in a 1-$\lambda$ planar microcavity formed by two AlGaAs/GaAs distributed Bragg reflectors (DBRs). The QD layer is located in the middle of the GaAs cavity between the DBRs, while the top (bottom) of the cavity is p-doped (n-doped), forming a p-i-n-i-n diode structure around the QDs. The sample is etched in two areas to two different depths so ohmic contact can be made directly to the doped regions of the cavity; we probe a QD in the un-etched area. A bias voltage can be applied across the diode to deterministically charge the QD with a single electron that tunnels in from the n-doped region \cite{warburton_optical_2000}. All measurements described here were performed using a bias in the center of the voltage range where the QD is negatively charged.
	
	A resonant laser is coupled into the waveguide mode of the microcavity through the cleaved edge of the sample. This allows discrimination between the waveguide mode that contains the excitation laser, and the Fabry-Perot mode that contains the QD fluorescence \cite{muller_resonance_2007,chen_resonance_2017}. This method has an advantage over schemes that use polarization to discriminate between fluorescence and laser scattering \cite{vamivakas_spin-resolved_2009,kuhlmann_dark-field_2013} because it provides full polarization freedom to the measurement. The resonant laser is used to coherently excite the QD and cause fluorescence. Scanning that laser over the resonances and recording the fluorescence intensity allows measurement of the excitation spectrum of the QD and reveals the direction and magnitude of any resonance shifts.
	
	\begin{figure*}[!hbt]
		\centering
		\includegraphics[keepaspectratio=true]{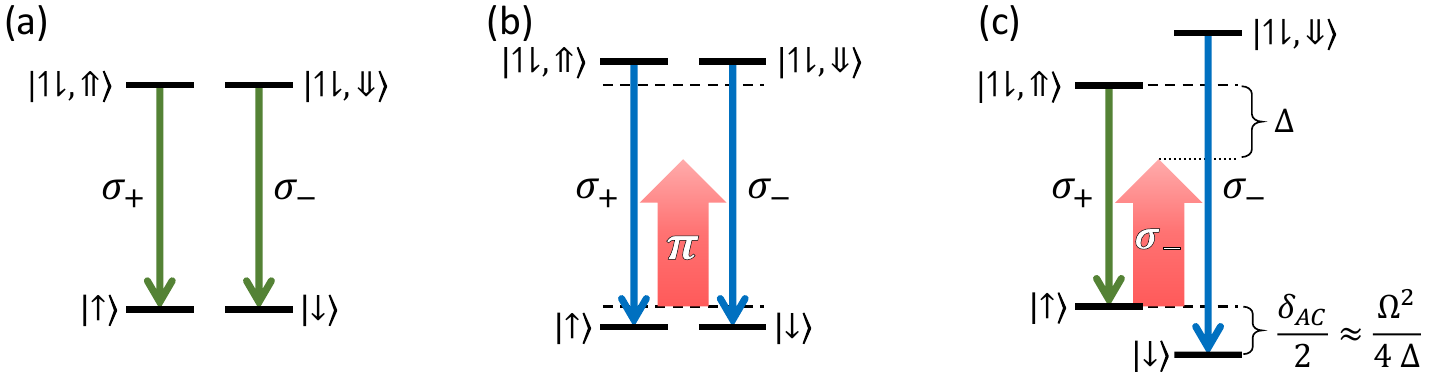}
		\caption{ \label{fig:EnergyStructure}
		Schematic of charged quantum dot energy levels and allowed transitions for (a) no AC Stark laser, (b) a red-detuned, linearly polarized AC Stark laser, and (c) a red-detuned, $\sigma_-$ polarized AC Stark laser. $\Delta$ is the detuning of the AC Stark laser from the QD resonance, $\Omega$ is the Rabi frequency of the AC Stark laser, and $\delta_{\mathrm{AC}}$ is the induced AC Stark shift of the $\sigma_-$ transition.}
	\end{figure*}
	
	A second laser used to cause the AC Stark effect is incident normal to the sample surface, anti-parallel to the growth direction. Unless otherwise noted, this AC Stark laser is far red-detuned from the QD transitions so as to minimize population of the excited states; blue-detuning causes incoherent excitation of the excited states via phonon coupling. A combination of polarizer and wave plates provide control over the polarization state of the AC Stark laser, which is used to selectively address one of the QD transitions. The spectrum of the laser is cleaned using a volume Bragg grating (VBG) to reflect only the strong laser line, which reduces the intensity of the amplified spontaneous emission (ASE) at the frequency of the QD transitions. The laser is then injected into the collection path via transmission through another VBG that is oriented to reflect the QD fluorescence into the measurement spectrometer. There is scattering of the AC Stark laser from the sample that must be attenuated so it does not overwhelm the weak QD fluorescence on the detector of the spectrometer. Even though the laser is spectrally distinct from the fluorescence because of the large red-detuning, the scattering is still strong enough that without any attenuation the diffuse reflections within the spectrometer would cause a large noise background. The same VBG that reflects the QD fluorescence allows most of the AC Stark laser scattering to transmit, reducing the intensity that reaches the spectrometer. A third VBG in the subsequent optical path allows the fluorescence to transmit and reflects the AC Stark laser line, further attenuating it. The result is that the laser scattering that reaches the spectrometer detector is attenuated enough that it can be spectrally discriminated from the QD fluorescence without requiring polarization discrimination. The ASE at the frequency of the QD is also attenuated sufficiently that even though it is not spectrally distinct from the fluorescence, it is still orders of magnitude weaker.
	
	The energy structure of a negatively charged QD has four energy levels as shown in Fig.~\ref{fig:EnergyStructure}(a). The two lower levels are the spin projection eigenstates of a single trapped electron. The two upper states are the spin eigenstates of a negative trion: two electrons in a singlet state and one heavy hole \cite{warburton_single_2013}. Without an external magnetic field, there are two transitions allowed by conservation of angular momentum: one between the spin-up states, which form the spin-up manifold; and one between the spin-down states, which form the spin-down manifold. These are spin-preserving cycling transitions in that a cycle of excitation and spontaneous emission returns the system to the same spin state in which it started. The cycling transitions are labeled $\sigma_+$ or $\sigma_-$ by the angular momentum gained by the QD during excitation: $\pm1$, measured along the growth direction. They correspond to circular polarization of the light.
	
	The AC Stark shift occurs when a far off-resonant laser is applied to a transition in the regime where the magnitude of the detuning, $\Delta = \omega_{\mathrm{laser}} - \omega_0$, is much greater than the Rabi frequency of the interaction, $\Omega$ \cite{flagg_optical_2015}. In that case, the excited states are only weakly populated to a degree $\frac{\Omega}{2\Delta}$, and both the ground and excited states involved in the transition are shifted in opposite directions by $\frac{\Omega^2}{4\Delta}$. The different power dependencies of the excited state population and the resonance shift mean that if $\Omega$ is large and $\Delta\gg\Omega$, then the shift can be large even while the excited population is small. The shifts of both states move the resonance frequency by $\delta_{\mathrm{AC}} = \frac{-\Omega^2}{2\Delta}$, which is the AC Stark shift we measure in the excitation spectra below. When the laser is red-detuned ($\Delta < 0$)  the AC Stark effect causes a blue-shift; when the laser is blue-detuned ($\Delta > 0$) it causes a red-shift. Figure \ref{fig:EnergyStructure}(b) shows the expected AC Stark shifts for a red-detuned, linearly polarized laser. Both spin manifolds are affected because  linear polarization is a superposition of $\sigma_+$ and $\sigma_-$ light. Figure \ref{fig:EnergyStructure}(c) shows the expected AC Stark shifts for a red-detuned, $\sigma_-$ polarized laser. Only one spin manifold is shifted by the AC Stark effect, and the magnitude of the shift is twice as large as that for linear polarization because the power of the laser is not split between two transitions.
	
	\begin{figure}[!hbt]
		\centering
		\includegraphics[keepaspectratio=true]{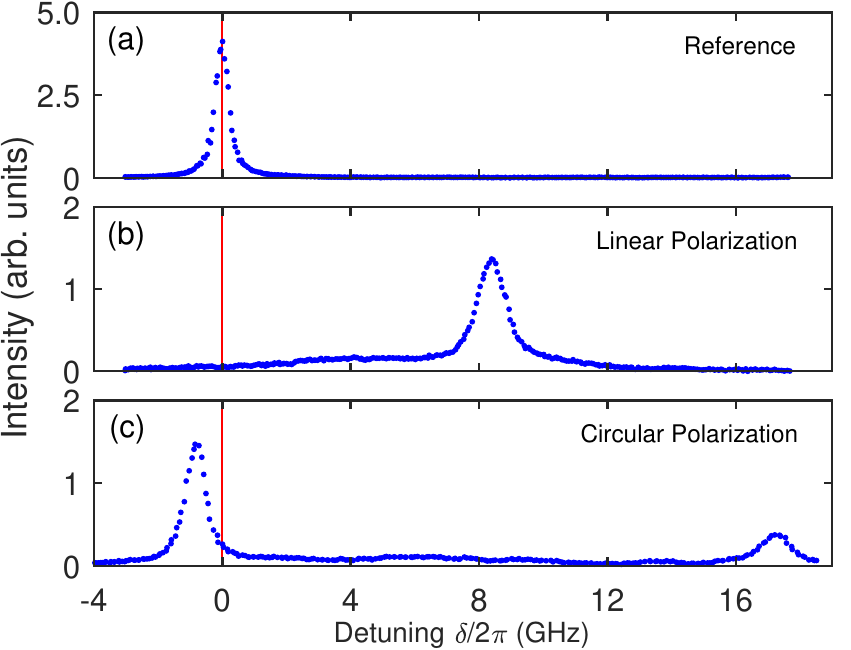}
		\caption{ \label{fig:ExcitationSpectra}
		Excitation spectra as functions of the resonant laser detuning, $\delta$. (a) Reference spectrum with no AC Stark laser. (b) Spectrum with linearly polarized ($\pi_y$) AC Stark laser. (c) Spectrum with circularly polarized ($\sigma_-$) AC Stark laser. For these measurements, the detection polarization was linear ($\pi_x$), the detuning of the AC Stark laser was $\Delta/2\pi = -1000$ GHz, and the power was 2.38 mW.}
	\end{figure}
	
	We measured the resonant excitation spectrum of the QD without the AC Stark laser to obtain a reference against which to compare the spectra under the influence of the AC Stark laser. The reference spectrum is shown in Fig.~\ref{fig:ExcitationSpectra}(a) and zero detuning ($\delta = 0$) is defined as the center frequency of the peak. There are two degenerate transitions in the reference spectrum, one for each spin manifold. We then measured excitation spectra while applying a linearly or circularly polarized AC Stark laser with a red-detuning of $\Delta/2\pi = -1000$ GHz and a power of 2.38 mW. The linear polarization shifts both transitions equally so they remain degenerate, as shown in Fig.~\ref{fig:ExcitationSpectra}(b). The circular polarization shifts one transition to higher energy by a large amount while the other transition remains near the reference frequency, as shown in Fig.~\ref{fig:ExcitationSpectra}(c). The lower-energy transition is actually red-shifted slightly due to interactions with the partially polarized nuclear spin ensemble of the atoms comprising the QD. This is called the Overhauser shift and is explained in greater detail below.
	
\begin{figure}[!b]
		\centering
		\includegraphics[keepaspectratio=true]{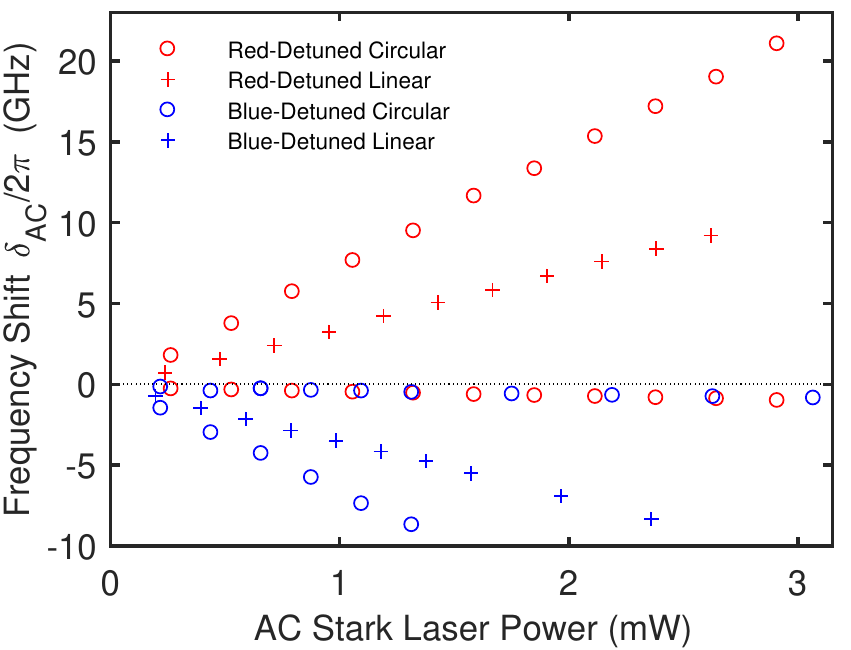}
		\caption{ \label{fig:FreqShifts}
		Frequency shifts of excitation spectra peaks as functions of the AC Stark laser power for red- or blue-detuning ($\Delta/2\pi = \pm1000$ GHz), and linear or circular polarization.}
	\end{figure}

	Figure \ref{fig:FreqShifts} shows the center frequencies of the peaks in the excitation spectra as functions of the AC Stark laser power. The frequency shifts for both linear and circular polarization can be seen for both red- and blue-detuning of the AC Stark laser. All the shifts are linear in the power of the AC Stark laser, as expected, because $\Omega\propto |\vec{E}|$ and $\Omega^2 \propto$ power. For linear polarization and red-detuning of the AC Stark laser, both transitions are shifted to higher frequency by the same amount; for blue-detuning both are shifted to lower frequency. For circular polarization and red-detuning, one transition is shifted to higher frequency by the combination of AC Stark effect and Overhauser field, while the other is shifted to lower frequency by the Overhauser field alone. For blue-detuning, both transitions are shifted to lower frequency, one by just the Overhauser field (as with red-detuning) and one by the combination of the AC Stark effect and the Overhauser field whose shifts are in opposition. Thus a blue-detuned AC Stark laser does not shift the affected spin manifold by as much as a red-detuned AC Stark laser. Regardless of the sign of the detuning of the AC Stark laser, the Overhauser shifts are in the same direction by the same amount. That is conclusive evidence that the observed shift is not due directly to the AC Stark effect, whose polarity would change sign with the detuning. The direction of the shift is consistent with the Overhauser effect expected due to dynamic nuclear polarization caused by electron spin pumping.
	
	Figure \ref{fig:ExcitationSpectra}(c) shows the excitation spectrum with a $\sigma_-$ AC Stark laser applied. In this case the $\sigma_-$ ($\sigma_+$) transition is blue- (red-) shifted from the reference frequency. The difference in the peak heights seen here means that the trapped electron has a non-zero time-averaged spin polarization, which increases fluorescence from the $\sigma_+$ transition. This implies that the electron spin is being pumped into the spin-up manifold. The exact mechanism of this spin pumping is not known at this time, but will be the subject of future investigations. We note that it is likely not due to the forbidden transition downward from the $\ket{\uparrow\downarrow,\Downarrow}$ trion state being weakly populated by the AC Stark laser, as the magnitude of the Overhauser shift seen in Fig.~\ref{fig:FreqShifts} is the same for both red- and blue-detuning. If this forbidden transition were the source of spin pumping we would expect to see a larger Overhauser shift with blue-detuning due to the incoherent excitation of the trion state via phonon coupling, as mentioned earlier. One possibility for the spin pumping mechanism is a process similar to that in Ref.~\cite{chekhovich_pumping_2010} where excitation of a forbidden transition is accompanied by a simultaneous spin flip of the electron and one of the nuclear spins. In this case the forbidden transition might be off-resonantly excited by the strong AC Stark laser. This process would result in the trapped electron spin being pumped into the spin-up manifold, consistent with the difference in the peak heights of Fig.~\ref{fig:ExcitationSpectra}(c).
	
	The time-averaged electron spin polarization is transferred to the nuclear spins by the contact hyperfine interaction \cite{eble_dynamic_2006}. The result is that the nuclear spin ensemble acquires a non-zero time- and ensemble-averaged polarization in the +z direction: $\braket{I_z} > 0$. The polarization of the nuclear spins causes an Overhauser shift of the electron and trion spin-states, which shifts the frequencies of the two optical transitions. The spin-down electron state $\ket{\downarrow}$ is shifted to lower energies, while the spin-down trion state $\ket{\uparrow\downarrow,\Downarrow}$ is shifted to higher energies, leading to an overall blue-shift of the $\sigma_-$ transition. The spin-up states shift in the opposite directions, leading to an overall red-shift of the $\sigma_+$ transition. Note that the direction of the Overhauser shifts depends only on the polarization of the AC Stark laser---not its detuning---whereas the AC Stark shifts depend on both the detuning and polarization of the laser. With a red-detuned, circularly polarized AC Stark laser, the Overhauser shift of the higher-frequency transition is in the same direction as the AC Stark shift, while the lower-frequency transition is shifted to lower frequency. With a blue-detuned AC Stark laser of the same polarization, the Overhauser shift is in the opposite direction from the AC Stark shift.  In both cases, the transition unaffected by the AC Stark shift is shifted to lower frequency by the Overhauser shift.
	
	\begin{figure}[!t]
		\centering
		\includegraphics[keepaspectratio=true]{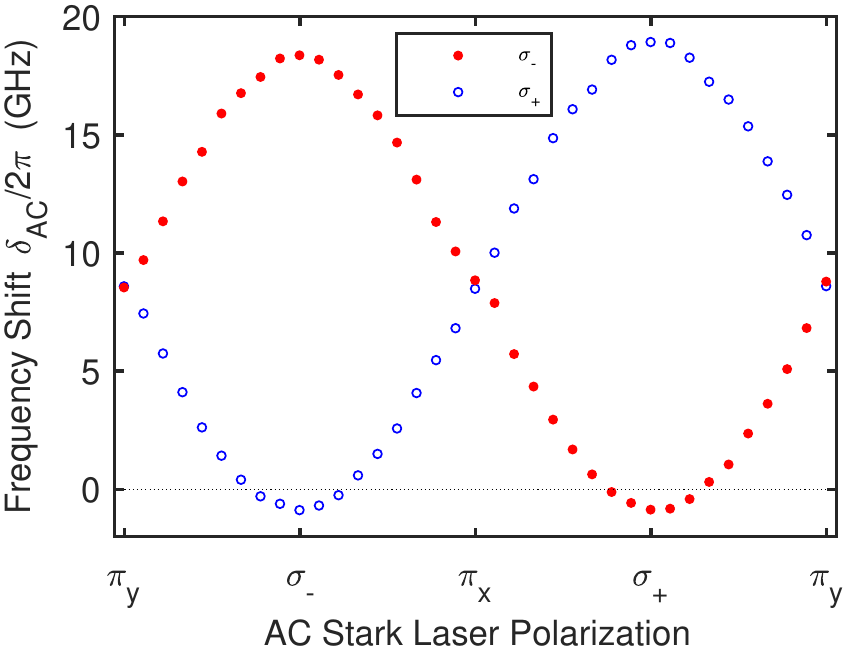}
		\caption{ \label{fig:FishPlot}
			Frequency shifts of excitation spectra peaks as a function of AC Stark polarization using orthogonal circular detection polarizations. The detuning of the AC Stark laser was $\Delta/2\pi = -1000$ GHz and the power was 2.6 mW for these measurements.}
	\end{figure}
	
	We can change the degree to which each transition shifts by changing the polarization of the AC Stark laser. Figure \ref{fig:FishPlot} shows the peak center frequencies as a function of the polarization of the AC Stark laser. For these measurements the laser power was 2.6 mW and it was red-detuned with $\Delta/2\pi = -1000$ GHz. For each AC Stark laser polarization, two excitation spectra were recorded, one with $\sigma_-$ detection and one with $\sigma_+$ detection. The detection polarization in turn determines which transition energy was measured. For linear laser polarization the transitions are nearly degenerate and blue-shifted from the reference frequency, as expected. As the laser becomes elliptically polarized the transitions split because the AC Stark effect influences them differently, which also causes electron spin pumping. The polarization of the electron spin causes dynamic nuclear polarization, and the Overhauser shift occurs. This is most obvious for circular polarization where the splitting is largest. In this case the lower-frequency transition is red-shifted by the Overhauser effect to below the reference frequency.
	
	We have demonstrated that spin-selective AC Stark shifts can be induced in a charged quantum dot using a strong, circularly polarized, far-detuned laser. We achieved a maximum splitting of approximately 20 GHz between the two spin-preserving transitions, which is large compared to the optical linewidth of about 1 GHz. The magnitude of the shift is in practice limited only by the amount of laser power that can be delivered to the QD while still discriminating the scattering from the fluorescence. We have optimized the efficiency of the optical path and the focus of the laser beam to maximize the shift achieved in this particular realization. The laser causing the shifts is red-detuned far enough that on its own it does not produce sufficient fluorescence to be measured above the noise background. However, the observed Overhauser shift implies that the AC Stark laser causes pumping of the electron spin. The electron spin polarization in turn causes nuclear polarization and an Overhauser field along the growth direction.
	
	The AC Stark shift achieved here is in principle sufficient to change the selection rules from the Voigt configuration of a small in-plane magnetic field to a pseudo-Faraday configuration with spin-preserving cycling transitions \cite{flagg_optical_2015}. The capability to rapidly and coherently reconfigure the energy structure and polarization selection rules of a charged QD enables a number of novel measurements. For example, it would allow measurements of electron and nuclear spin properties in zero magnetic field. It is also possible that the enhanced spin state splitting caused by the effect could protect the electron spin from magnetically induced dephasing \cite{golter_protecting_2014}. In a larger context, the capability of switching between Voigt and Faraday configurations will allow the quantum optics research community to investigate more complex control and manipulation sequences for single-qubit operations.
	
	\begin{acknowledgments}
	This research was supported by the U.S. Department of Energy, Office of Basic Energy Sciences, Division of Materials Sciences and Engineering under Award DE-SC0016848.
	\end{acknowledgments}

	%

	\bibliography{ACStark_references}
	
\end{document}